\documentclass[aps,prl,twocolumn,10pt,nofootinbib,showpacs]{revtex4-1} 
\usepackage{amsmath}
\usepackage{graphicx}
\usepackage{float}
\usepackage{amsfonts}
\usepackage{braket}
\usepackage{natbib}
\usepackage{amssymb}
\usepackage{fancyhdr}
\usepackage[utf8]{inputenc}
\usepackage[colorlinks=true,citecolor=blue,linkcolor=blue]{hyperref}
\usepackage{dsfont}
\usepackage{subfigure}

\newcommand{\LR}[1]{\left(#1\right)}

\begin{document}

\title{Adaptive hybrid optimal quantum control for imprecisely characterized systems}
\author{D. J. Egger}
\author{F. K. Wilhelm}
\affiliation{Theoretical Physics, Universität des Saarlandes, D-66123 Saarbrücken, Germany}

\begin{abstract}
Optimal quantum control theory carries a huge promise for quantum technology. Its experimental application, however, is often hindered by imprecise knowledge of the input variables, the  quantum system's parameters. We show how to overcome this by Adaptive Hybrid Optimal Control (Ad-HOC). This protocol combines open-  and closed-loop optimal control by first performing a gradient search towards a near-optimal control pulse and then  an experimental fidelity estimation with a gradient-free method. For typical settings in solid-state quantum information processing, Ad-HOC enhances gate fidelities by an order of magnitude, making optimal control theory applicable and useful.
\end{abstract}

\maketitle

\begin{figure*}[htbp!] \centering
 \includegraphics[width=0.8\textwidth]{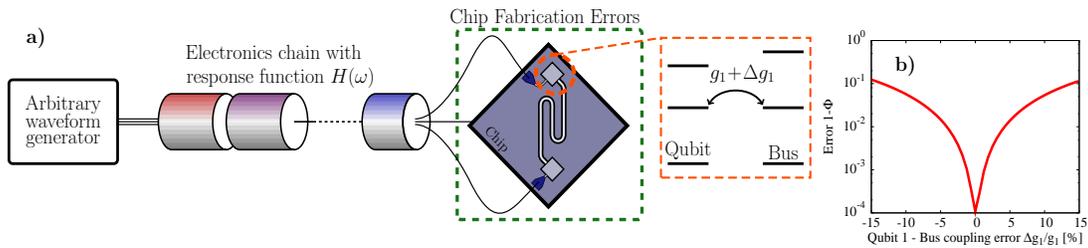}
 \caption{(color online). a) Sketch of quantum control experiment. The unit generating the control pulses, typically an arbitrary waveform generator (AWG), at room temperature generates the control pulses that are sent through the control transfer chain (sketched as the chain of cylinders) to finally reach the quantum system, often cooled to less than a Kelvin. Error sources are in the parameters modeling the ``chip'', the electronics and the calibration of the control signals. b) Rapid degradation of a 99.99\% fidelity CZ gate between two qubits coupled via a bus assuming only an error on the $g_1$.
 \label{Fig:Fig1}}
\end{figure*}

The last decades have seen the transformation of quantum theory from a mere description of nature to a tool in research and applications, prominently in quantum information processing \cite{Nielsen_and_Chuang_book}, spectroscopy, sensing, and metrology \cite{Boto00}. Quantum control describes the science of shaping the time evolution of quantum systems in a potentially useful way \cite{Rice00,Brumer03}. Control parameters typically are parameters of an external field parameterized in a technologically appropriate way, e.g., into a quantum logic gate \cite{Spoerl07}, into a higher coherence in NMR \cite{Khaneja_JMR_172_296305,Glaser98,Borneman10}, or into states important for sensing \cite{Lapert12}. While analytically accessible only in highly specialized cases, these pulse shapes can in many cases be found using the powerful mathematical technique of Optimal Control Theory (OCT); by solving a Schr\"odinger or master equation iteratively, a pulse shape producing the desired time evolution can be found \cite{Khaneja_JMR_172_296305}. This results in complex pulses that are used in a wide variety of cases such as controlling the cooperative effects of driving and dissipation \cite{Schmidt_PRL_107_130404}, to control non-integrable quantum many-body \cite{Doria_PRL_106_190501} and many electron \cite{Castro_PRL_109_153603} systems, generating matter-wave entanglement \cite{Brucker_NatPhys_7_608, Jager_PRA_88_035601} and quantum information devices \cite{Nigmatullin_NJP_11_105032,Montangero_PRL_99_170501,Motzoi_PRL_103_110501}. These pulses are designed based on the best available knowledge of the system. This can be insufficient for two reasons i) In many cases, the underlying model cannot be solved with sufficient precision as in the case of many-body systems
\cite{Castro_PRL_109_153603} ii) in quantum systems that are engineered or when a human-made apparatus is a key part of the setup, parameters need to be measured with precision compatible with the control task at hand \cite{Egger_SUST_27_014001}, which is often not possible. This necessity to precisely know the underlying model strongly limits harvesting the benefits of optimal control in complex quantum systems.

In this Letter we solve this problem with a hybrid open/closed-loop optimal control method called Adaptation by hybrid optimal control (Ad-HOC). It combines a model based gradient search and the model free Nelder-Mead (NM) algorithm \cite{Nelder_CompJ_7_308}. Ad-HOC is designed to overcome shortcomings of the assumed physical model \cite{Hellgren_PRA_88_013414}, errors on the controls and inaccurate knowledge of the parameters.  We demonstrate this approach along two tasks: We first show that pulses can be optimized using only feedback from the experiment. We then show the efficiency of the hybrid method for the example of two superconducting qubits \cite{Clark_Nature_453_1031}.

Model-free calibration was pioneered for state transfers in chemical reactions \cite{Judson_PRL_68_1500} using genetic algorithms and was implemented for state transfer in optical lattices in \cite{Rosi_PRA_88_021601}. The many successes of this method as well as improvements can be found in \cite{Brif_NJP_12_075008}. We in turn optimize gates, i.e. transfers of a full basis of Hilbert space over a short distance in the control landscape, a task for which we found NM to be 1.5 orders of magnitude faster. The NM algorithm has been used in tuning dynamical decoupling sequences in \cite{Biercuk_Nature_458_996} and is part of the Chopped RAndom Basis
(CRAB) optimal control scheme \cite{Doria_PRL_106_190501} without initial gradient search. The closed-loop part of Ad-HOC has been experimentally demonstrated on a CZ gate between two coupled superconducting qubits \cite{Kelly_in_prep} and enabled the high gate fidelities in \cite{Barends_arXiv14024848}.

{\em Problem setting:} Delicate engineering of controlled quantum systems, in particular the need to isolate quantum systems from their environments, makes quantum control setups very complex. Such an experiment, sketched in Fig. \ref{Fig:Fig1} is made of the system to be controlled and the unit (the AWG) producing the control pulses. The pulses are brought from the latter to the former by a chain of electronic or optical components referred to as \emph{control transfer chain}. We assume that this chain and the AWG have a sufficiently large bandwidth to manipulate the system in the required way. In this setup, four different mechanisms will degrade the fidelity of an OCT designed pulse. i) Parameter estimation: The quantum system to be controlled is modeled by a drift and control Hamiltonians $\hat H=\hat H_\text{d}+\sum_i u_i(t)\hat H_{\text{c},i}$ with $\boldsymbol u(t)$ the control fields to be shaped. Imprecise characterization of parameters entering the drift $\hat H_\text{d}$ and controls $\hat H_{\text{c},i}$ will degrade fidelity. ii) Improper characterization of the control transfer chain's distortion of the pulses \cite{Jager_PRA_88_035601,Motzoi_PRA_84_022307}. iii) Signal calibration: in practice the control unit generates an electrical signal or laser impulse which is related to $\boldsymbol u(t)$. Imprecisions in this relation, e.g. a constant offset, generate errors on the controls. iv) Effects that are not taken into account in $\hat H$. Among many examples are other idling components of a complex quantum system such as a quantum processor, spurious two level systems (TLS) in Josephson Junctions, as well as slow non-Markovian noise. Errors in parameters and controls could be addressed in viewing the experiment as part of an ensemble and then using broadband control \cite{Owrutsky_PRA_86_022315,Khani_PRA_85_022306}. This typically leads to cumbersome pulses since high-order Lie brackets have to be generated by the compensating pulse \cite{Li_PRA_73_030302}.  Instead with Ad-HOC the pulses are suited to the {\em single} yet {\em uncertain} physical system at hand, thus avoiding complexity based on a simpler task.

\begin{figure}[htbp!]\centering
 \includegraphics[width=0.35\textwidth]{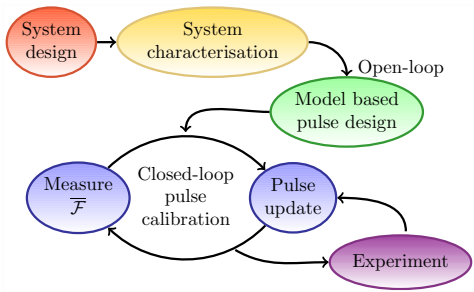}
 \caption{(color online). Ad-HOC protocol: The physical system and surrounding control and measurement apparatus are designed taking control problems into consideration. For instance the AWG has to have a sufficient bandwidth for the desired control task. The system is then characterized with the best possible precision. Using the resulting parameters the control pulses are created. These are then fine tuned to the system using closed-loop OCT. The pulses are then ready to be used in the experiment and can be recalibrated at a later time to account for drift. \label{Fig:Fig2} }
\end{figure}

{\em Proposed method:} In order to address imperfections of the model, the control loop can be  closed by using the experiment as feed-back to calibrate the control pulses. An initial gradient search \cite{Floether_NJP_14_073023}, e.g. done with the the GRadient Ascent Pulse Engineering (GRAPE) algorithm \cite{Khaneja_JMR_172_296305}, of the optimal pulse, taking into account constraints on the controls as well as robustness is performed with the best reasonably achievable (to be quantified) model of the system. This gives control pulses that yield high fidelity on the model but perform sub-optimally in the real system. As long as the model is a reasonably good approximation of the physical system, these pulses will still lie close to the optimal point in the control landscape. A set of similar pulses (with model parameters drawn from the error bars of the initial characterization of the system) are sent to the experiment and their performance measured. The pulses are then updated and the procedure is iterated until either a target performance is reached or convergence halts. Measuring pulse performance is time consuming, thus we chose the NM algorithm \cite{Nelder_CompJ_7_308}. It is robust and typically only evaluates 1-2 pulses per iteration. Once the calibration is done, the pulses can be used. At a latter time a few pulse calibration iterations correct for drifts in parameter values and experiments can resume. The Ad-HOC protocol is illustrated in Fig. \ref{Fig:Fig2}. Note that the precise experimental parameters are never identified. Ad-HOC hinges on an efficient method to experimentally estimate the performance index. Here, the performance index is the process fidelity which can be estimated using Randomized Benchmarking (RB) \cite{Magesan_PRL_109_080505,Chow_PRL_102_090502,Magesan_PRL_106_180504}. Other than standard process tomography,  it is significantly faster to obtain and minimizes the impact of state preparation and measurement errors. RB yields the average fidelity 
\begin{align} \label{Eqn:agf}
 \overline{\mathcal{F}}=\int\mathrm{d}\hat U\braket{\psi|\hat U^\dagger \hat U_\text{t}^\dagger\Lambda(\hat U\ket{\psi}\!\!\bra{\psi}\hat U^\dagger)\hat U_\text{t}\hat U|\psi},
\end{align}
estimating how well the channel $\Lambda$ implements the target Clifford gate $\hat U_\text{t}$. As shown in \cite{Kelly_in_prep}, RB is well-adapted to fast experimentation and catches a variety of practical errors of different scales.

In summary, the gradient search approaches a favorable control over a large distance based on theory and simulation whilst the closed-loop design, done on the experiment, takes into account all experimental details \cite{Judson_PRL_68_1500}.

\begin{figure}[htbp!]\centering
 \includegraphics[width=0.48\textwidth]{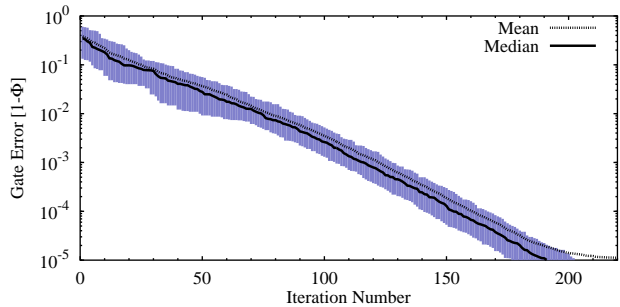}
 \caption{(color online). Convergence during optimization of random gates. 100 pulses were optimized each for a different realization of the random TLS. The target fidelity of $1-10^{-5}$ is reached rapidly as indicated by the median. The shaded area includes 68\% of all runs centered around the median.  \label{Fig:Fig3_a} }
\end{figure}

{\em Closed loop  optimization:} To show that a pulse can be optimized based only on its performance index we consider random gate synthesis. Inside a black box is a TLS in which the drift and control Hamiltonians are both random Hermitian matrices. The black box input is a pulse and the output its fidelity. The target is a random unitary matrix. Figure \ref{Fig:Fig3_a} shows the mean and median error as function of iteration for 100 different realizations of the TLS (see supplementary material for details \cite{Supplement}). The convergence is consistent with an exponential decrease of the error as a function of the number of steps. It is important to recognize in Fig. \ref{Fig:Fig3_a} that while demonstrating the power of the closed-loop part of Ad-HOC it also highlights that closed-loop control alone needs a large number of steps for a rather elementary control task. Going down this convergence curve with gradient search drastically reduces the number of steps to about 50 per order of magnitude error reduction.

{\em Numerical demonstration for a realistic setting:} To demonstrate hybrid optimal control in a more complicated yet realistic and genuine system, we choose to create a CZ gate between two superconducting qubits in the qubit-bus-qubit system \cite{Mariantoni_Sci_334_6165, Galiautdinov_PRA_85_042321}. These systems are well described by the typical setup of Fig. \ref{Fig:Fig1}. The qubit-bus-qubit Hamiltonian is modeled by
\begin{align}\notag
 \hat H=\sum\limits_i \delta_i(t)\hat \sigma^+_i\hat\sigma_i^- +\Delta_i\ket{2}_{i\,i}\!\bra{2}+\frac{g_i}{2}\LR{\hat\sigma_i^+\hat a+\hat\sigma_i^-\hat a^\dagger}\,.
\end{align}
The control $\delta_i(t)$ is the $i^\text{th}$ qubit-bus detuning. Their coupling strength is $g_i$. $\Delta_i$ is the qubit's non-linearity. $\hat\sigma_i^+$ and $\hat a^\dagger$ respectively create an excitation in qubit $i$ and the bus. This system is particularly vulnerable to errors on the controls and parameters \cite{Egger_SUST_27_014001}. For instance Fig. \ref{Fig:Fig1}\textcolor{blue}{b} shows the fidelity loss due to a small error on $g_1$. 5\% imprecision increases the error by two orders of magnitude. In fact, albeit the initial numerical optimization leading to a pulse that is first-order insensitive to errors, the second derivative is large, making this an example that is specifically unforgiving to model uncertainty and the ideal case for showing Ad-HOC's performance.

\begin{figure}\centering
 \includegraphics[width=0.8\columnwidth]{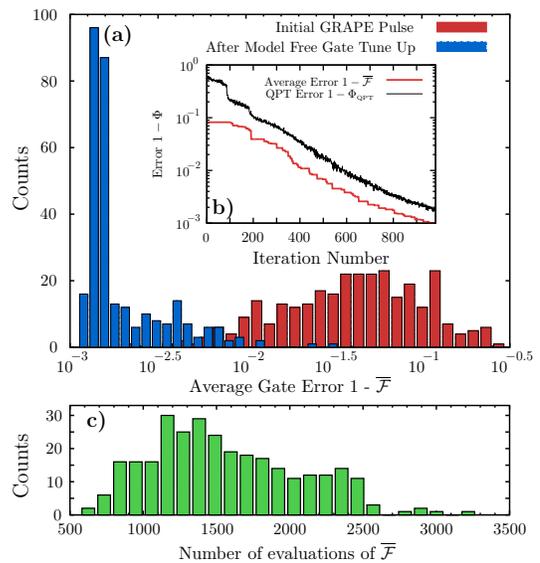}
 \caption{(color online). Debugging procedure for parameter errors, control transfer chain errors and control DC offset errors. The error of the initial pulse was minimized using a gradient search down to machine precision. The pulses are then calibrated to a specific realization of the system. a) Histograms for 300 system realizations. The red histograms show the fidelity of the initial uncalibrated numerical pulse. The blue histograms show the improvement in average gate fidelity after running Ad-HOC. b) Gate errors as function of the calibration algorithms iteration number. c) Histogram of the number evaluations of $\overline{\mathcal{F}}$ needed to calibrate the pulse, i.e. to take the red histograms to the blue ones. \label{Fig:Fig4}}
\end{figure}

First, a gradient search optimizes down to machine precision the error of a CZ gate using the quantum process fidelity $\Phi=|\text{Tr}\{\hat U_\text{cz}^\dagger\, \hat U[\delta_1,\delta_2]\}|^2/d^2$. $\Phi$ measures the overlap between the ideal CZ gate $\hat U_\text{cz}$ and the gate implemented by the controls $\delta_i$. $d$ is the dimension of the Hilbert space. GRAPE optimizes $\Phi$ by slicing time into intervals across which the controls are constant, i.e. $\delta_i(t)\to\{\delta_{ij}\}$. It then searches in the direction of steepest $\partial\Phi/\partial\delta_{ij}$ which can be computed analytically \cite{Machnes_PRA_84_022305}. Next, the model parameters $g_i$ and $\Delta_i$, as well as the standard deviation of the transfer chain's impulse response are promoted to random variables following Gaussian statistics with variances reflecting the precision of actual parameter estimations \cite{Lucero_NatPhys_8_719}. Additionally, random calibration offsets are introduced on the pulses. The difference between the new and old optimal controls is five times smaller than between the initial GRAPE guess and the resulting optimal control (see supplementary material \cite{Supplement}). We then compute the average gate fidelity $\overline{\mathcal{F}}$ for many different realizations of the system, see the red histograms in Fig. \ref{Fig:Fig4}\textcolor{blue}{a}. As expected the fidelities are nowhere close to optimal ranging between 99\% and 68\%, clearly insufficient for quantum computing \cite{Fowler_PRA_86_032324}. Finally each instance is reoptimized using the closed loop part of Ad-HOC, i.e., a pulse for that specific parameter set is found. For each realization,  Ad-HOC increased the fidelity by more than an order of magnitude, as seen by the blue histograms in Fig. \ref{Fig:Fig4}\textcolor{blue}{a}. Fig. \ref{Fig:Fig4}\textcolor{blue}{b} shows a typical decrease in error during the closed loop optimization. As $\overline{\mathcal{F}}$ is being maximized, $\Phi$, computed for comparison, also increases. The corresponding number of required evaluations of $\overline{\mathcal{F}}$ for each realization  is shown in Fig. \ref{Fig:Fig4}\textcolor{blue}{c}. 

\begin{figure}[htbp!]\centering
 \includegraphics[width=0.8\columnwidth]{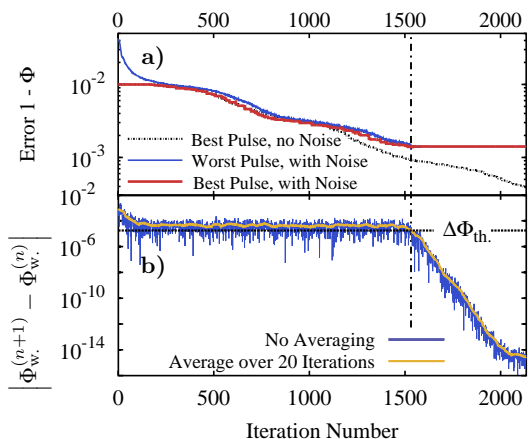}
 \caption{(color online). a) Convergence speed of a single optimization comparing the cases when a depolarizing channel adds noise and when the optimization is noiseless. b) Difference in fidelity of the worst point in the simplex between subsequent iterations in a noisy optimization. As long as, on average, this difference is greater than the noise level, the optimization continues. \label{Fig:Fig5}}
\end{figure}

\emph{ Robustness:} Unlike pure open-loop techniques, the robustness of Ad-Hoc is limited by the reliability of fidelity estimation. In the previous examples the sampling of the integral in Eq. (\ref{Eqn:agf}) introduces noise into the fidelity estimation. Noise would also be present in an experiment but for different reasons. Here is further investigated the effect of noise on convergence. We consider the fidelity $\Phi$ which can be computed without introducing noise. A noiseless run of closed-loop optimization is compared to one with noise artificially added by a depolarizing channel \cite{Nielsen_and_Chuang_book}. Both optimizations are shown in Fig. \ref{Fig:Fig5}, they converge at the same speed until the noisy case halts. This termination results from the increase in fidelity, averaged over several iterations, being smaller than the noise threshold $\Delta\Phi_\text{th.}$ (see supplementary material \cite{Supplement}). The calibration protocol can no longer determine if the changes made to the pulses improve $\Phi$ and halts. This is illustrated in Fig. \ref{Fig:Fig5}\textcolor{blue}{b} showing the difference between successive iterations of fidelity of the worst pulse $\Phi_\text{w.}$ in the NM simplex.

In conclusion we have proposed Adaptive Hybrid Optimal Control (Ad-HOC), a protocol for overcoming model imperfection and incompleteness afflicting the design of control pulses for quantum systems. The protocol is efficient and can be applied to almost arbitrary quantum control experiments as it can be used with any fidelity measure that captures the essence of the desired time evolution. We showed that noise does not affect convergence speed but rather the terminal fidelity. Therefore higher fidelity can be gained by increasing the estimation precision. The closed-loop part of Ad-HOC has been demonstrated in \cite{Kelly_in_prep}.

We thank J.M. Martinis for insisting that optimal control will not be applied without calibration, M. Biercuk for pointing us to the NM algorithm and J. Kelly and R. Barends for pointing out the speediness of Randomized Benchmarking. This work was supported by the EU through SCALEQIT and QUAINT as well as funded by the Office of the Director of National Intelligence (ODNI), Intelligence Advanced Research Projects Activity (IARPA), through the Army Research Office.

\bibliographystyle{apsrev4-1}
\bibliography{AdHOC.bib}{}

\newpage
\onecolumngrid
\appendix

\section{Supplementary material}

In this supplementary material we give some details on the optimization of control pulses for random two level systems. We also show that a few parameter pulse can be calibrated extremely quickly. A few extra details on the qubit-bus-qubit system are given. Finally we show that the calibration does not have to be constrained to the use of randomized benchmarking. To do this we calibrate a CZ pulse using a fidelity measure tailored to the CZ gate.

\subsection{Modeling of the Control Transfer Chain}
The transfer chain between the quantum system and the arbitrary waveform generator (AWG) can be taken into account in the optimization \cite{Jager_PRA_88_035601,Motzoi_PRA_84_022307}. However, improper characterization of it will degrade pulse performance. Here we describe how control transfer chains can be modeled and how output signals from the AWG relate to the control fields $\boldsymbol u(t)$ used in the Hamiltonian. The voltages $\boldsymbol V(t)$ produced by the AWG are not identical to the functions $\boldsymbol u(t)$. Instead they are related through a calibration curve $ \mathcal{\boldsymbol C}$. Furthermore the impulse response of the transfer chain $h$ between the AWG and the experiment can distort the pulses. Thus whilst the AWG produces $\boldsymbol V(t)$ the quantum system actually receives
\begin{equation}
 \boldsymbol u(t)=\int\limits_{0}^t {\rm d \tau} \LR{\mathcal{\boldsymbol C} \circ \boldsymbol V}\!(t-\tau) \,  h(\tau).
\label{eq:lineartransfer}
\end{equation}
This can be taken into account using the methodology of Ref.~\cite{Motzoi_PRA_84_022307}  if  $\mathcal{\boldsymbol C}$ and $h$ are precisely known. Practically, these functions as well as the linearity of the signal transfer stipulated in Eq. (\ref{eq:lineartransfer}) are hard to verify with the needed precision. Whereas errors in parameters of the system can be addressed using broadband control \cite{Skinner_JMR_163_8, Owrutsky_PRA_86_022315, Khani_PRA_85_022306}, we know of no such approach for uncertain transfer functions.

\subsection{Control of Random Two Level Systems}
\begin{figure}[htbp!] \centering
 \includegraphics[width=0.8\textwidth]{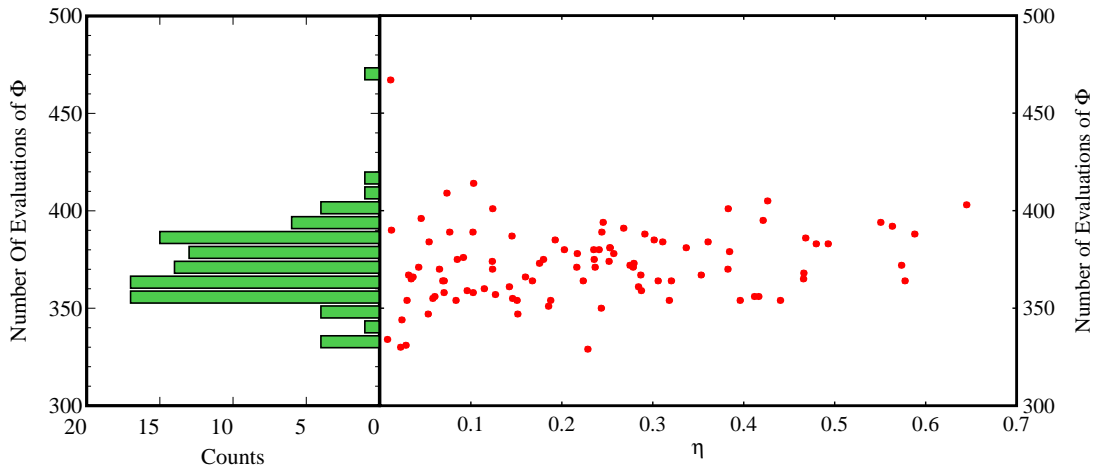}
 \caption{ Number of evaluations of the fidelity as function of the biggest relevant norm $\eta$ defined in Eq. (\ref{Eqn:Supp_Eta}). In this data two points with particularly small $\eta$ failed to converge and were excluded from the data. \label{Fig:Supp_MatNorm}}
\end{figure}

To investigate the performance of the model free calibration we apply it to the control of random two level systems. The Hamiltonians are
\begin{align} \notag
 \hat H(t)=\begin{pmatrix} H^{(\text{d})}_1 & H^{(\text{d})}_2+iH^{(\text{d})}_3 \\ H^{(\text{d})}_2-iH^{(\text{d})}_3 & H^{(\text{d})}_4 \end{pmatrix} +u(t)\begin{pmatrix}H^{(\text{c})}_1 & H^{(\text{c})}_2+iH^{(\text{c})}_3 \\ H^{(\text{c})}_2-iH^{(\text{c})}_3 & H^{(\text{c})}_4\end{pmatrix}.
\end{align}
The random variable $H^{(\text{x})}_i\in\mathds{R}$ are uniformly distributed in $[-0.5,0.5]$. For each realization of the drift $\hat H^{(\text{d})}$ and control $\hat H^{(\text{c})}$ a target unitary matrix $\hat U_\text{rand}$, chosen randomly is given. For each realization we seek a different control $u(t)$ to optimize the fidelity $\Phi=|{\rm Tr}\{\hat U_\text{rand}^\dagger \hat U\}|^2/4$. A histogram of the number of runs required to reach $1-10^{-5}$ fidelity is shown in Fig. \ref{Fig:Supp_MatNorm}. The median and mean fidelity as function of number of iterations is shown in the main text. Instances that converged poorly can be attributed to realizations that are hard to control in the given time, as the commutator between $\hat{H}^{\rm d}$ and $\hat{H}^{\rm c}$ turns out to be too small. To confirm this statement we plot the number of times $\Phi$ was evaluated as function of the smallest relevant matrix norm, defined as
\begin{align}
 \eta=\max&\left\{ \left\|\left[\hat H^{(\text{d})},\hat H^{(\text{c})}\right]\right\|,\left\|\left[\hat H^{(\text{d})}, [\hat H^{(\text{d})},\hat H^{(\text{c})}]\right]\right\|,\left\|\left[\hat H^{(\text{c})}, [\hat H^{(\text{d})},\hat H^{(\text{c})}]\right]\right\|\right\}\,. \label{Eqn:Supp_Eta}
\end{align}
$\|\cdot\|$ is the $\max$ norm. The smaller $\eta$ is, the harder the system is to control. This is reflected in Fig. \ref{Fig:Supp_MatNorm}. Overall for controllable systems the number of evaluations of $\Phi$ is low especially since the starting point for the optimization was the null control $u(t)=0~\forall~t$. At very small values of $\eta$ the system tends to be uncontrollable and some target gates cannot be reached. Two bad instances were removed from the data. These had very small commutator norms and would have required a much longer gate time.

\subsection{Few Parameter Pulse Closed-Loop Optimization Example: Finding DRAG \label{Sec:NM_PiPulse}}
\begin{figure}[htbp!] \centering 
 \subfigure[$~$Initial pulse sequence \label{Fig:NM_PiPulses_I}]{\includegraphics[width=0.35\columnwidth]{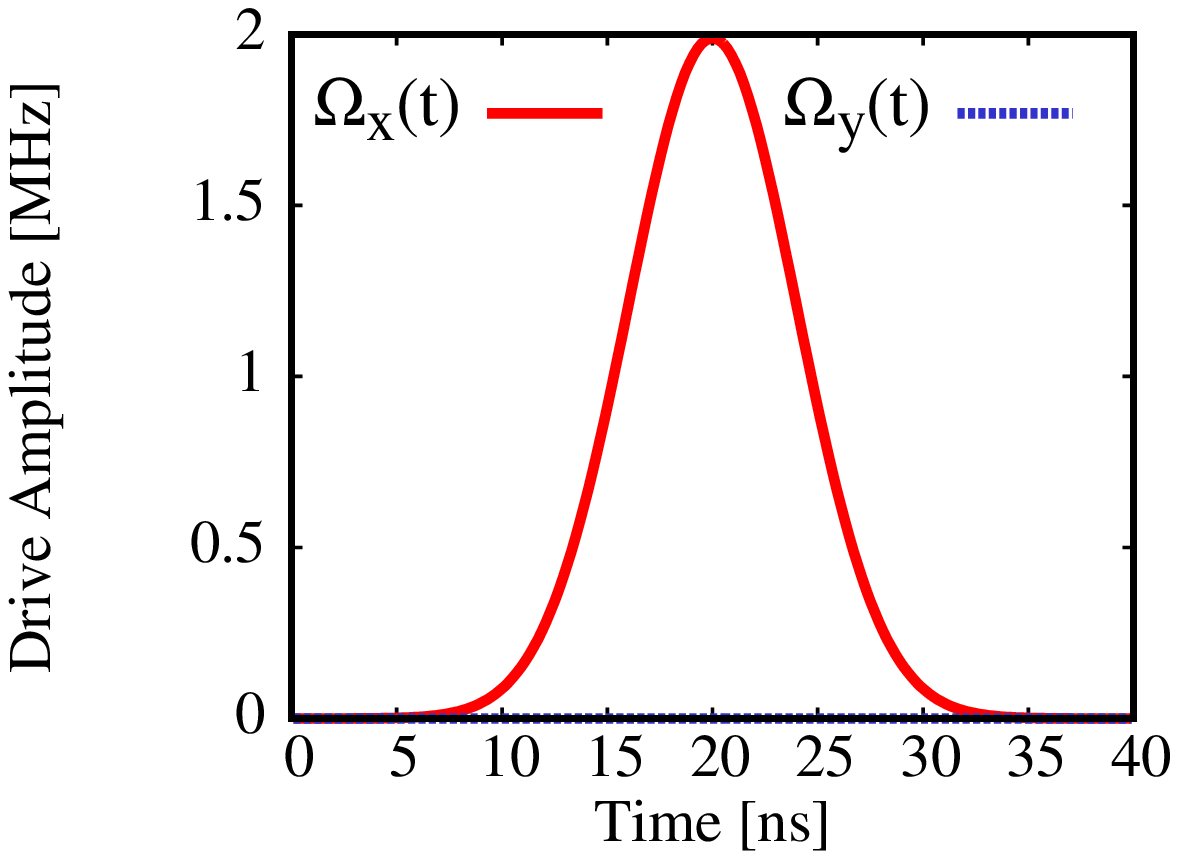}}
 \subfigure[$~$Evolution of $\ket{1}\bra{1}$ before optimization. \label{Fig:NM_PiPulses_PI}]{\includegraphics[width=0.35\columnwidth]{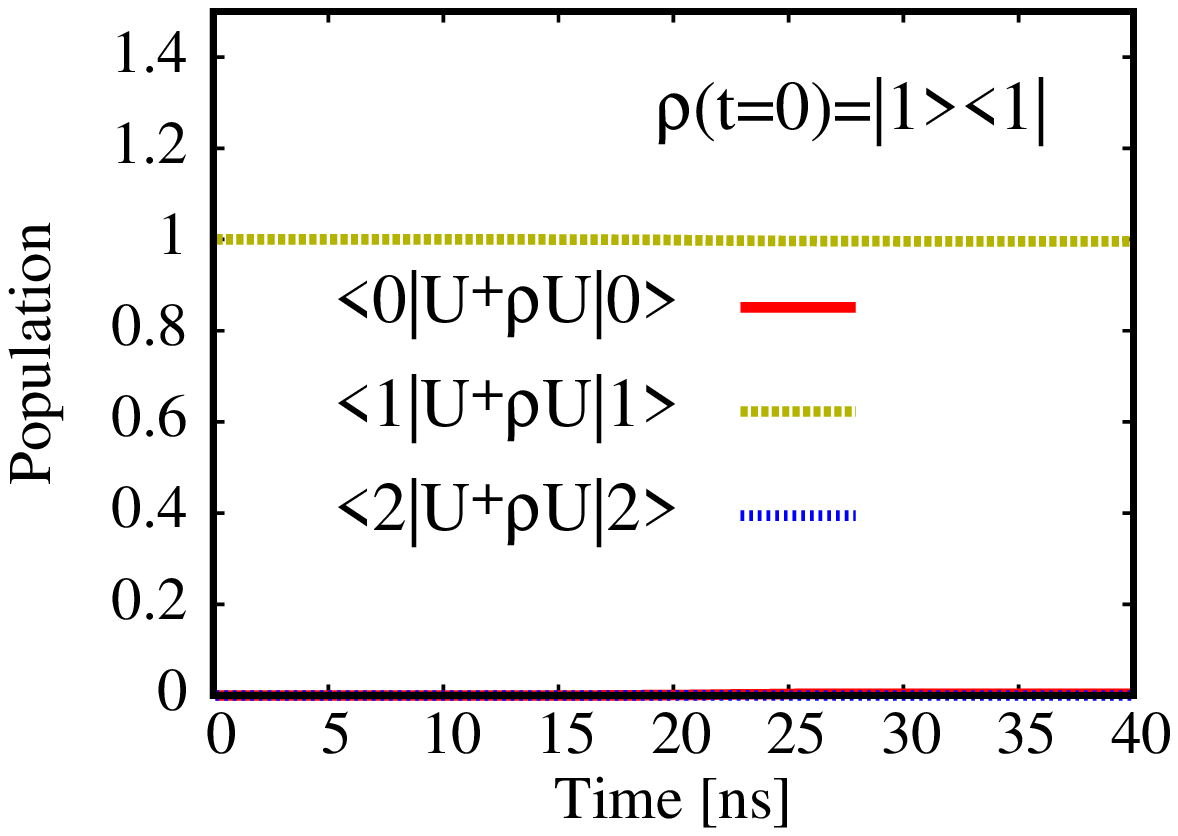}}
 \subfigure[$~$Final pulse sequence \label{Fig:NM_PiPulses_F}]{\includegraphics[width=0.35\columnwidth]{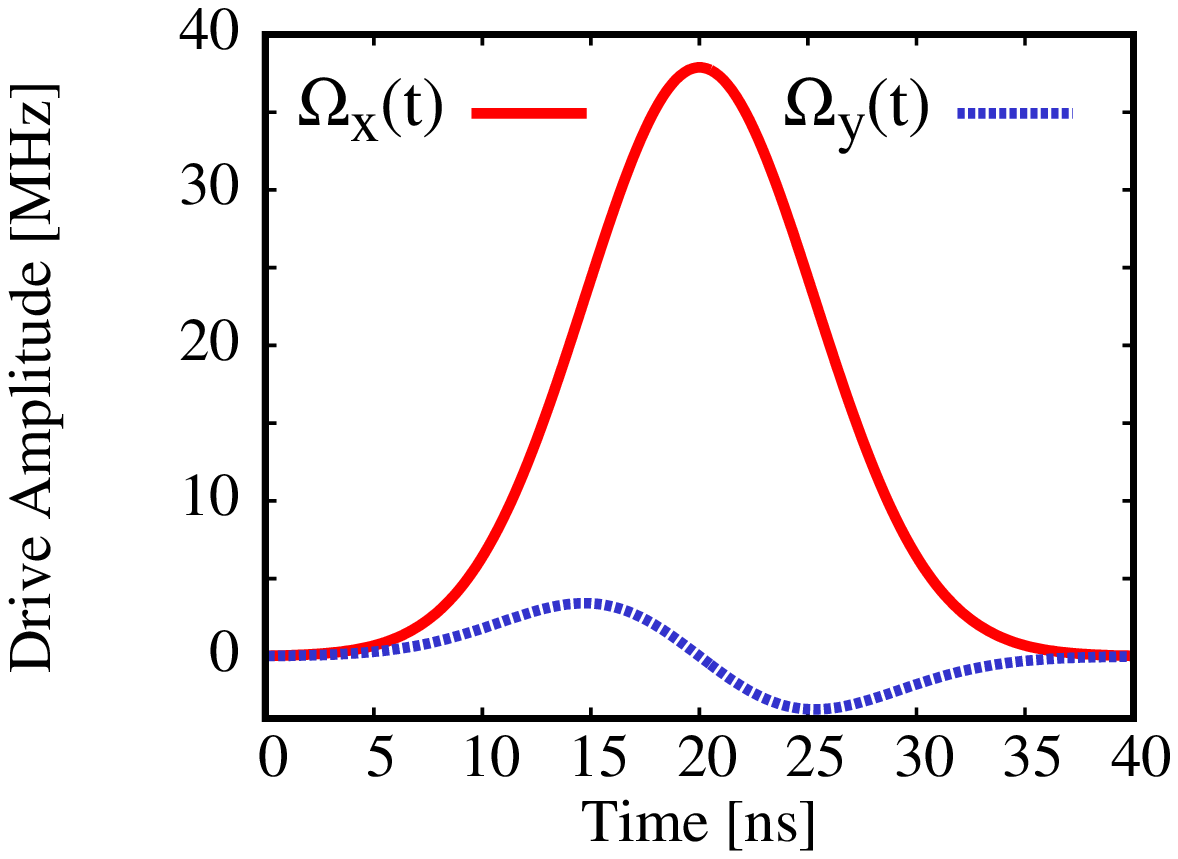}}
 \subfigure[$~$Evolution of $\ket{1}\bra{1}$ after optimization. \label{Fig:NM_PiPulses_PF}]{\includegraphics[width=0.35\columnwidth]{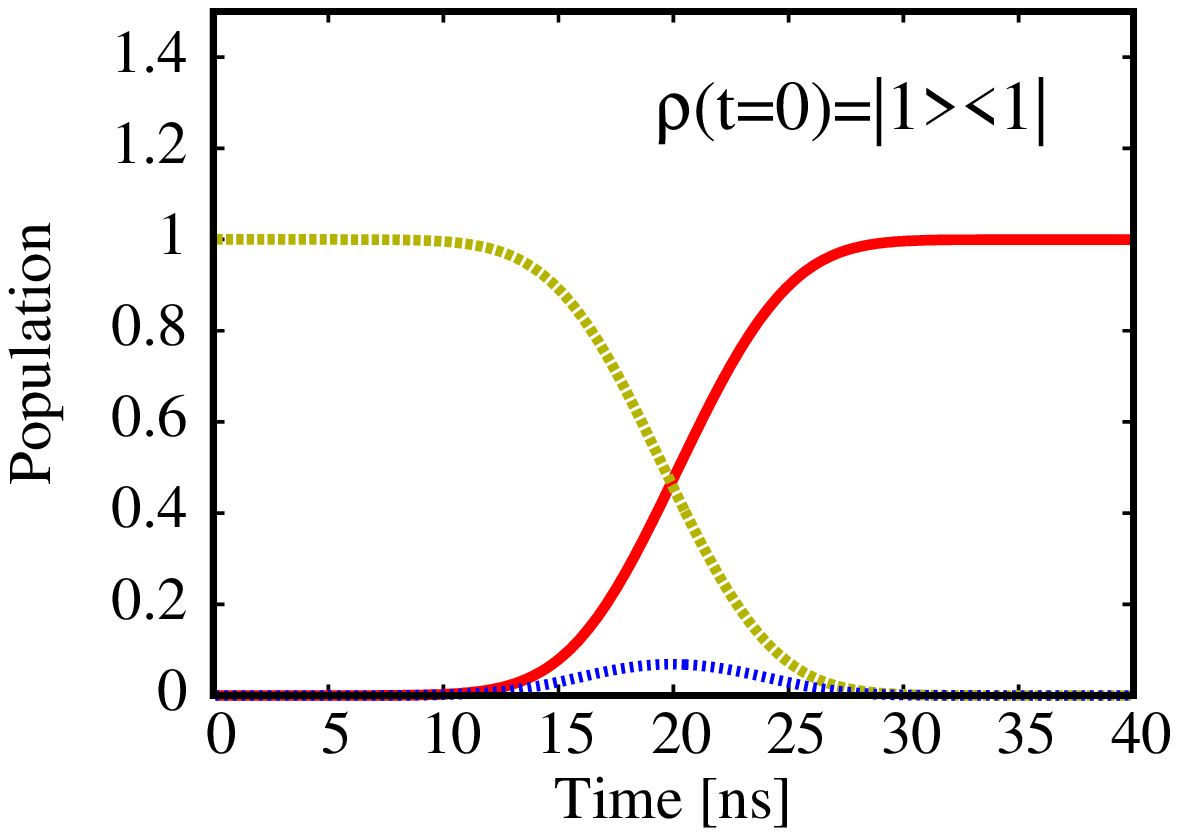}}
 \caption{ Control pulse of a weakly non-linear three level system. The target time evolution is an $\hat X$ gate. A bad initial pulse fails to produce the desired time evolution. \subref{Fig:NM_PiPulses_I} and \subref{Fig:NM_PiPulses_PI} show the initial pulse and the corresponding population evolution when starting with $\ket{1}\bra{1}$. The final pulse, found after few iterations, produces an $\hat X$ gate while minimizing leakage, see \subref{Fig:NM_PiPulses_F} and \subref{Fig:NM_PiPulses_PF} respectively for pulse and population as function of time.  \label{Fig:NM_PiPulses}}
\end{figure}

\begin{figure}[htbp!] \centering
 \includegraphics[width=0.4\columnwidth]{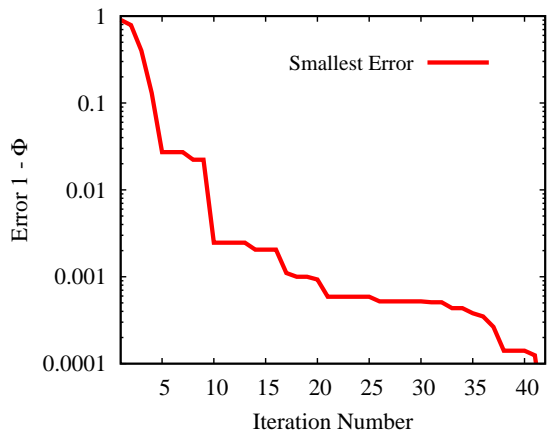}
 \caption{Improvement of the fidelity as function of iteration number for the pulses of Fig. \ref{Fig:NM_PiPulses}. \label{Fig:Error_Iter_Pi}}
\end{figure}

In a two level system an $\hat X$ gate can be implemented by applying a pulse of area $\pi$ on the $\Omega_x$ component of the driving field. However, when a third level is present this is no longer sufficient and DRAG pulses are needed \cite{Motzoi_PRL_103_110501, Gambetta_PRA_83_012308}. Generally, full characterization of the third level - its anharmonicity and coupling ratio, is an extra characterization step that can be avoided using Ad-HOC. The Hamiltonian for an anharmonic three level system, driven on resonance and in the frame rotating at the frequency of the drive field is
\begin{align} \notag
 \hat H = \begin{pmatrix} 0 & 0 & 0 \\ 0 & 0 & 0 \\ 0 & 0 & \Delta \end{pmatrix}+\frac{\Omega_x(t)}{2}\begin{pmatrix} 0 & 1 & 0 \\ 1 & 0 & \sqrt{2} \\ 0 & \sqrt{2} & 0 \end{pmatrix}+i\frac{\Omega_y(t)}{2}\begin{pmatrix} 0 & -1 & 0 \\ 1 & 0 & -\sqrt{2} \\ 0 & \sqrt{2} & 0 \end{pmatrix}\,.
\end{align}
$\Delta$ is the anharmonicity also called qubit non-linearity. To drive the $0\leftrightarrow1$ transition without driving $1\leftrightarrow2$ the $\Omega_y$ quadrature has to be set to the derivative of $\Omega_x(t)$ scaled by $-1/2\Delta$. To show that few parameter pulses can be quickly calibrated, we assume that the anharmonicity is not known and that the initial pulse is a Gaussian with the wrong area
\begin{align} \notag
 \Omega_{x,\text{initial}}(t)=A\exp\left\{-\frac{t^2}{2\sigma^2}\right\}~~~\Omega_{y,\text{initial}}(t)=0.
\end{align}
Here $A$ and $\sigma$ are chosen at random. The calibration protocol has to find the correct values for $A$, $\sigma$ and $\Delta$ such that the time evolution is
\begin{align} \notag
 \hat U_\text{target} = \begin{pmatrix} 0 & 1 & 0 \\ 1 & 0 & 0 \\ 0 & 0 & e^{i\varphi} \end{pmatrix}
\end{align}
An example of the pulses are shown in Fig. \ref{Fig:NM_PiPulses}. Figures \ref{Fig:NM_PiPulses_I} and \ref{Fig:NM_PiPulses_F} respectively show the pulses before and after the optimization which took only 76 evaluations of the fidelity function $\Phi=|{\rm Tr}\{\hat X^\dagger \hat U\}|^2/9$. As can be seen by Fig. \ref{Fig:NM_PiPulses_PI} the initial pulse is unable to drive any transitions since the amplitude of the pulse is too weak. The fidelity as function of iteration number for these pulses is shown in Fig. \ref{Fig:Error_Iter_Pi}. Closed-loop optimization quickly finds the optimal pulse.

\newpage

\subsection{Application of Ad-HOC to the Qubit-Bus-Qubit Architecture}
Using the rotating wave approximation, the Hamiltonian of the qubit-bus-qubit architecture in the frame rotating at the frequency of the bus is 
\begin{align}\label{Eqn:Hqbq}
 \hat H=\sum\limits_{i=1}^2 \delta_i(t)\hat \sigma^+_i\hat\sigma_i^- +\Delta_i\ket{2}_{i\,i}\!\bra{2}+\frac{g_i}{2}\LR{\hat\sigma_i^+\hat a+\hat\sigma_i^-\hat a^\dagger}=\hat H_\text{d}+\sum\limits_{i=1}^2\delta_i(t)\hat H_{\text{c},i}\,.
\end{align}
The qubits are modeled as anharmonic three level systems with creation operators $\hat\sigma_i^+$ and non-linearity $\Delta_i$. They are coupled to the bus with strength $g_i$. The bus creation operator is $\hat a^\dagger$. The parameters $g_i$ and $\Delta_i$ need to be measured resulting in parameter errors. We approximate the control transfer chain by convoluting the pulse with a Gaussian function of standard deviation $\sigma_\text{filt}$. The open loop optimization is done by gradient search with the quasi-Newton method BFGS \cite{Floether_NJP_14_073023}. In experiments the effect of the electronics is not as simple but can be taken into account by deconvoluting the controls with a measured transfer function. Imprecisions in this measurement further introduce errors. We model this by promoting the standard deviation of our Gaussian convolution function to a random variable following a Gaussian distribution. Lastly calibration errors between the output of the AWG and the qubit frequency can cause the qubit to over or under estimate the qubit-bus resonance point \cite{Egger_SUST_27_014001}. Since it is this resonance point which is most crucial to the gate we model it by a DC offset of $\Delta\omega_{\text{b},i}$ on the controls. This introduces the error term
\begin{align}\notag
 \hat H_\text{err}=\sum\limits_{i=1}^2\Delta\omega_{\text{b},i}\hat\sigma^+_i\hat\sigma^-_i
\end{align}
in the Hamiltonian (\ref{Eqn:Hqbq}), it now reads $\hat H'=\hat H+\hat H_\text{err}$. Furthermore, in practice $\Delta\omega_{\text{b},i}$ is not perfectly known and is therefore promoted to a random variable. In summary the parameters used in the model are given in Tab. \ref{Tab:Param}. The imprecision reflects current experiments \cite{Lucero_NatPhys_8_719}.

\begin{table}[htbp!]
 \begin{center}
  \caption{Parameters used in the model. The coupling strength $g$ is given as function of the Qubit-Bus swap time by $(2T_\text{swap})^{-1}$. The imprecision is given relative to the parameter it refers to. When promoting the parameters to random variables this imprecision serves as standard deviation. When performing the closed loop optimization, the AWG voltage calibration $\Delta\omega_{\text{b},i}$ is chosen randomly with mean zero and standard deviation of 0.1\% of the bus frequency. System realizations with unphysical parameters are discarded, e.g. $\sigma_\text{filt}$ cannot be smaller than zero. \label{Tab:Param}}
  \begin{tabular}{l c c c c r} \hline\hline
   Name & Symbol & Qubit 1 & Qubit 2 & unit & Imprecision  \\ \hline
   Qubit-Bus Swap Time & $T_{\text{swap},i}$ & 12.6 & 9.2 & ns & -\\
   Qubit-Bus Coupling Strength & $g_i$ & 40 & 54 & MHz & 4.0\% \\
   Qubit non-linearity & $\Delta_i$ & -59 & -71 & MHz & 4.0\%\\ \hline
   Bus Frequency & $\omega_\text{b}$ & \multicolumn{2}{c}{6.1} & GHz & 0.1\% \\ 
   Convolution function error & $\sigma_\text{filt}$ & \multicolumn{2}{c}{1} & ns & 10.0\% \\ \hline\hline
  \end{tabular}
 \end{center}
\end{table}

\subsubsection{Control Distance Characterization}
\begin{figure}\centering
  \includegraphics[width=0.45\columnwidth]{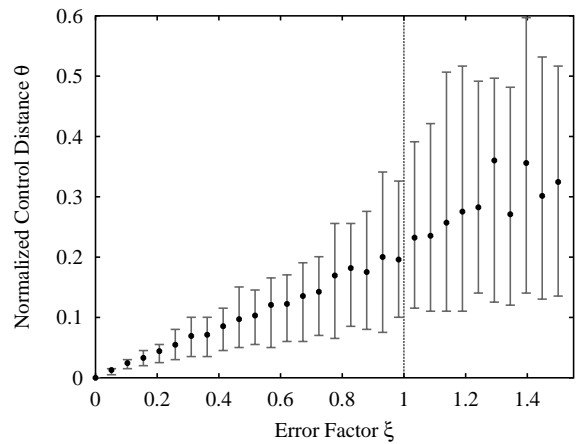}
  \caption{Study of the distance between the model optimal controls and physical system optimal controls. The distance $\theta$, defined in Eq. \ref{Eqn:ctrl_dist} is measured as function of the size of the error, controlled by the scaling parameter $\xi$. This plot shows that the model and system optimal controls still lie close to one another. \label{Fig:ctrl_dist}}
\end{figure}

The gradient search done on the model allows for a large distance to be covered in the control landscape. Generally the initial guess $\delta_{i,\text{init}}(t)$ is far away from the model optimal controls $\delta_{i,\text{mod}}(t)$. The closer the model is to the physical system, the less iterations are needed by the closed-loop part of Ad-HOC to reach the physical system optimal controls $\delta_{i,\text{sys}}(t)$. We quantify the difference in distance between the model optimal and system optimal controls by
\begin{align} \label{Eqn:ctrl_dist}
 \theta(\xi)=\frac{\displaystyle\sum\limits_{i=1}^2\int\limits_0^T\left\vert\delta_{i,\text{sys}}(t,\xi)-\delta_{i,\text{mod}}(t)\right\vert{\rm d}t}{\displaystyle\sum\limits_{i=1}^2\int\limits_0^T\left\vert\delta_{i,\text{mod}}(t)-\delta_{i,\text{init}}(t)\right\vert{\rm d}t}\,.
\end{align}
$\xi$ is a parameter that we use to control the size of the errors. Note that here only the system optimal controls $\delta_{i,\text{sys}}$ depend on $\xi$. When generating the physical system realizations, according to the parameters in Tab. \ref{Tab:Param}, the standard deviation, i.e. the imprecision, is scaled by $\xi$. This allows us to control the difference between the model and system. When $\xi=0$ the model and the system are identical. The results for the qubit-bus-qubit system are shown in Fig. \ref{Fig:ctrl_dist}. The vertical dashed line at $\xi=1$ corresponds to achievable precisions in current experiments \cite{Lucero_NatPhys_8_719}. The fact that $\theta(\xi=1)\simeq 0.2$, indicates that the system and model optimal controls are a lot closer to each other than the model optimal controls and the GRAPE initial guess. The latter is an educated guess that brings both qubits on resonance with the bus.

\subsubsection{Additional Example}

\begin{figure}[htbp!]\centering
 \subfigure[$~$Before closed-loop part of Ad-HOC]{\includegraphics[width=0.47\columnwidth]{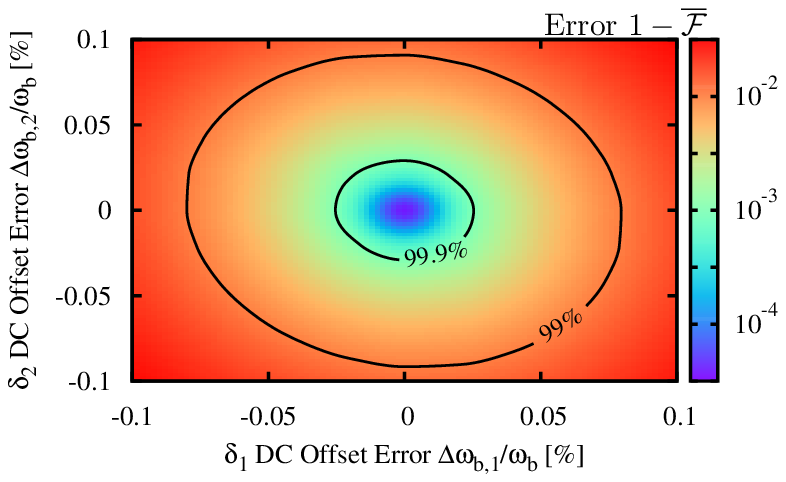}} \hfill
 \subfigure[$~$After closed-loop part of Ad-HOC]{\includegraphics[width=0.47\columnwidth]{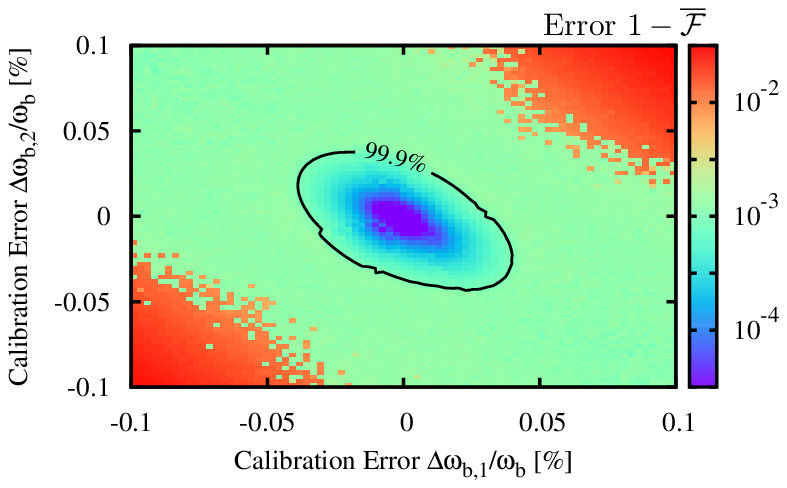}}
 \caption{Average error for the CZ gate as function of the DC offset error $\Delta\omega_{\text{b},i}$ introduced by miscalibrating the AWG's output to the qubit frequency. Ad-HOC greatly improves the fidelity of the pulse as can be seen by comparing Fig. \textcolor{blue}{a)} and \textcolor{blue}{b)}. The central region of high fidelity does not change since the target fidelity for the calibration protocol was 99.9\%. \label{Fig:Fig4b}}
\end{figure}

Additionally to the example in the main text, a further illustration of Ad-HOC's performance is shown in Fig. \ref{Fig:Fig4b}. In this case only the control DC offset error $\Delta\omega_{\text{b},i}$ was present. As can be seen in Fig. \ref{Fig:Fig4b} the fidelity has been increased over a wide range of possible $\Delta\omega_{\text{b},i}$'s. This shows how successful Ad-HOC is in dealing with errors on the controls introduced by $\hat H_\text{err}$.

\subsubsection{Optimization with Gate-Taylored Quality Parameters}

\begin{figure}[htbp!] \centering
 \includegraphics[width=0.4\textwidth]{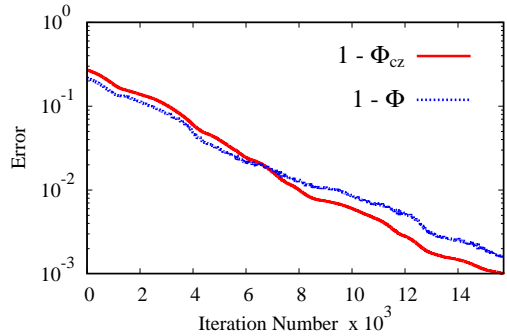}
 \caption{ Model free calibration with a gate tailored fidelity. At each iteration the gate overlap fidelity is also computed and, as can be seen, optimizing $\Phi_\text{cz}$ also optimizes $\Phi$. \label{Fig:Phi_Exp_Ex}}
\end{figure}

The main text emphasizes randomized benchmarking as the fidelity measure. This fidelity is applicable when the desired gate is a Clifford gate. Here is shown that a gate specific fidelity can also be used to calibrate the pulse. We illustrate this with the optimization of a CZ gate. We define the following fidelity measure
\begin{align} \notag
 \Phi_\text{cz}=\!\!\!\!\!\!\sum\limits_{ij\in\{01,10,11\}}\!\!\!\!\!\!\frac{\left| U_{ij,ij}\right|^2}{6} \left[1+(-1)^{ij} \cos \LR{\text{arg}\LR{U_{ij,ij}}}\right]. \label{Eqn:ExpPhi}
\end{align}
Here $U_{ij,ij}$ is the element of the time evolution operator mapping the state $\ket{ij}$ onto itself. The terms $| U_{ij,ij}|^2$ are the qubit populations after the pulse sequence for a specific input state. The argument of these terms can be found using Ramesy measurements. A gate that is unitary and optimizes $\Phi_\text{cz}$ has to be a good CZ gate. An example of this fidelity as function of the iteration number is shown in Fig. \ref{Fig:Phi_Exp_Ex}. The initial pulse was optimized by GRAPE up to $80\%$ fidelity using $\Phi=|{\rm Tr}\{\hat U_\text{cz}^\dagger \hat U\}|^2/d^2$. The remaining calibration was done with the model free part of Ad-HOC. The target fidelity was set to be $\Phi_\text{cz}=99.9\%$. It can be seen that the intrinsic gate fidelity follows closely.

\section{Noise Threshold}
Fig. \textcolor{blue}{5} of the main text shows that the closed-loop optimization stops when it can no longer tell, on average, if the changes in fidelity are due to noise or changes in the pulse parameters. The noisy fidelity used was
\begin{align}\notag
 \Phi_\text{noisy}=\frac{n}{md}+\frac{m-n}{m}\Phi
\end{align}
where $n$ is the number of times the channel depolarized out of a total of $m$ trials. $d$ is the Hilbert space dimension. Therefore the uniform probability of depolarization $p$ is estimated by $n/m$. In order for the closed loop part of Ad-HOC to converge it must be able to distinguish if a new pulse sequence is better or worse. This sets bounds on the amount of noise tolerated. Thus when, on average, an operation on the simplex improves $\Phi$ of the worst pulse by less than a threshold difference $\Delta\Phi_\text{th.}$ the optimization will not be able to improve the fidelity any longer because the experiment cannot distinguish fidelities sufficiently well. For this case, the threshold difference is
\begin{align}\notag
 \Delta\Phi_\text{th.}=\frac{d\Phi_\text{dep.}-\bar p-\sigma_p}{d(1-\bar p-\sigma_p)}-\frac{d\Phi_\text{dep.}-\bar p+\sigma_p}{d(1-\bar p+\sigma_p)}\,.
\end{align}
The estimation of $p$ is $\bar p\pm\sigma_p=n/m\pm1/12\sqrt{m}$. Here the factor 12 comes from the uniform distribution.

\end{document}